\documentclass[twocolumn,showpacs,aps,prd]{revtex4}

\usepackage{graphicx}
\usepackage{dcolumn}
\usepackage{amsmath}
\usepackage{epsfig}

\input ./babarsym

\newcommand{\BABARPubYear}    {13}
\newcommand{\BABARPubNumber}  {008}
\newcommand{\SLACPubNumber} {15835}

\def\figurebox#1#2#3{%
    \def\arg{#3}%
    \ifx\arg\empty
    {\hfill\vbox{\hsize#2\hrule\hbox to #2{\vrule\hfill\vbox to #1{\hsize#2\vfill}\vrule}\hrule}\hfill}%
    \else
    {\hfill\epsfbox{#3}\hfill}%
    \fi}

\begin{document}
\preprint{\babar-PUB-\BABARPubYear/\BABARPubNumber} 
\preprint{SLAC-PUB-\SLACPubNumber}

\begin{flushleft}
\babar-PUB-\BABARPubYear/\BABARPubNumber\\
SLAC-PUB-\SLACPubNumber\\
\end{flushleft}

\title{
{\large \bf
\boldmath
Evidence for the baryonic decay $\BDLL$
\unboldmath
} 
}

%% author list as of 03-Apr-2013 (339 authors)
%
\author{J.~P.~Lees}
\author{V.~Poireau}
\author{V.~Tisserand}
\affiliation{Laboratoire d'Annecy-le-Vieux de Physique des Particules (LAPP), Universit\'e de Savoie, CNRS/IN2P3,  F-74941 Annecy-Le-Vieux, France}
\author{E.~Grauges}
\affiliation{Universitat de Barcelona, Facultat de Fisica, Departament ECM, E-08028 Barcelona, Spain }
\author{A.~Palano$^{ab}$ }
\affiliation{INFN Sezione di Bari$^{a}$; Dipartimento di Fisica, Universit\`a di Bari$^{b}$, I-70126 Bari, Italy }
\author{G.~Eigen}
\author{B.~Stugu}
\affiliation{University of Bergen, Institute of Physics, N-5007 Bergen, Norway }
\author{D.~N.~Brown}
\author{L.~T.~Kerth}
\author{Yu.~G.~Kolomensky}
\author{M.~J.~Lee}
\author{G.~Lynch}
\affiliation{Lawrence Berkeley National Laboratory and University of California, Berkeley, California 94720, USA }
% ...
\collaboration{The \babar\ Collaboration}
\noaffiliation

\begin{abstract}
Evidence is presented for the baryonic \B meson decay \BDLL based on a data sample of $471 \times 10^6$ \BBbar pairs collected with the \babar~detector at the \pep2 asymmetric \epem collider located at the SLAC National Accelerator Laboratory. 
The branching fraction is determined to be $\BR(\BDLL) = (9.8^{+2.9}_{-2.6} \pm 1.9)\times 10^{-6}$, corresponding to a significance of $3.4$ standard deviations including systematic uncertainties. A search for the related baryonic \B meson decay \BDSL with $\Sigmaz\to\Lambda\g$ is performed and an upper limit $\BR(\BDSL+\BDLS) < 3.1 \times 10^{-5}$ is determined at $90\%$ confidence level.
\end{abstract}


\maketitle

\section{Introduction}
\label{sec:introduction}

Little is known about the mechanism of baryon production in weak decays or in the hadronization process. Baryons are produced in $(6.8 \pm 0.6)\%$ of all \B meson decays~\cite{ref:PDG}. Due to this large rate, $B$ meson decays can provide important information about baryon production.
Due to the low energy scale, perturbative quantum chromodynamics (QCD) cannot be applied to this process. Furthermore, latice QCD calculations are not available. The description of baryonic \B decays thus relies on phenomenological models.\newpage 
Pole models \cite{ref:Pole} are a common tool used in theoretical studies of hadronic decays.
Meson pole models predict an enhancement at low baryon-antibaryon masses.
In many three-body decays into a baryon, an antibaryon and a meson, the baryon-antibaryon pair can be described by a meson pole, i.e., the decay of
a virtual meson with a mass below threshold.
This leads to a steeply falling amplitude at the threshold of the baryon-antibaryon mass,
and explains the enhancement observed in decays such as
$\Bm\to\LC\antiproton\pim$~\cite{ref:StephBelle, ref:Steph}, $\Bm\to\pp\Km$~\cite{ref:ppk, ref:ppkBelle, ref:ppkLHCb}, and $\Bzb\to\Dz\pp$~\cite{ref:Dpp, ref:DppBelle}.

In addition to the meson pole models described above, there
are baryon pole models in which the initial state decays through the strong interaction into a pair of baryons. Then, one of these baryons decays via the weak interaction into a baryon and a meson. For such baryon pole models no enhancement at threshold in the dibaryon invariant mass is expected.

\begin{figure*}[!t]
 \centering
   \includegraphics[width=.33\textwidth]{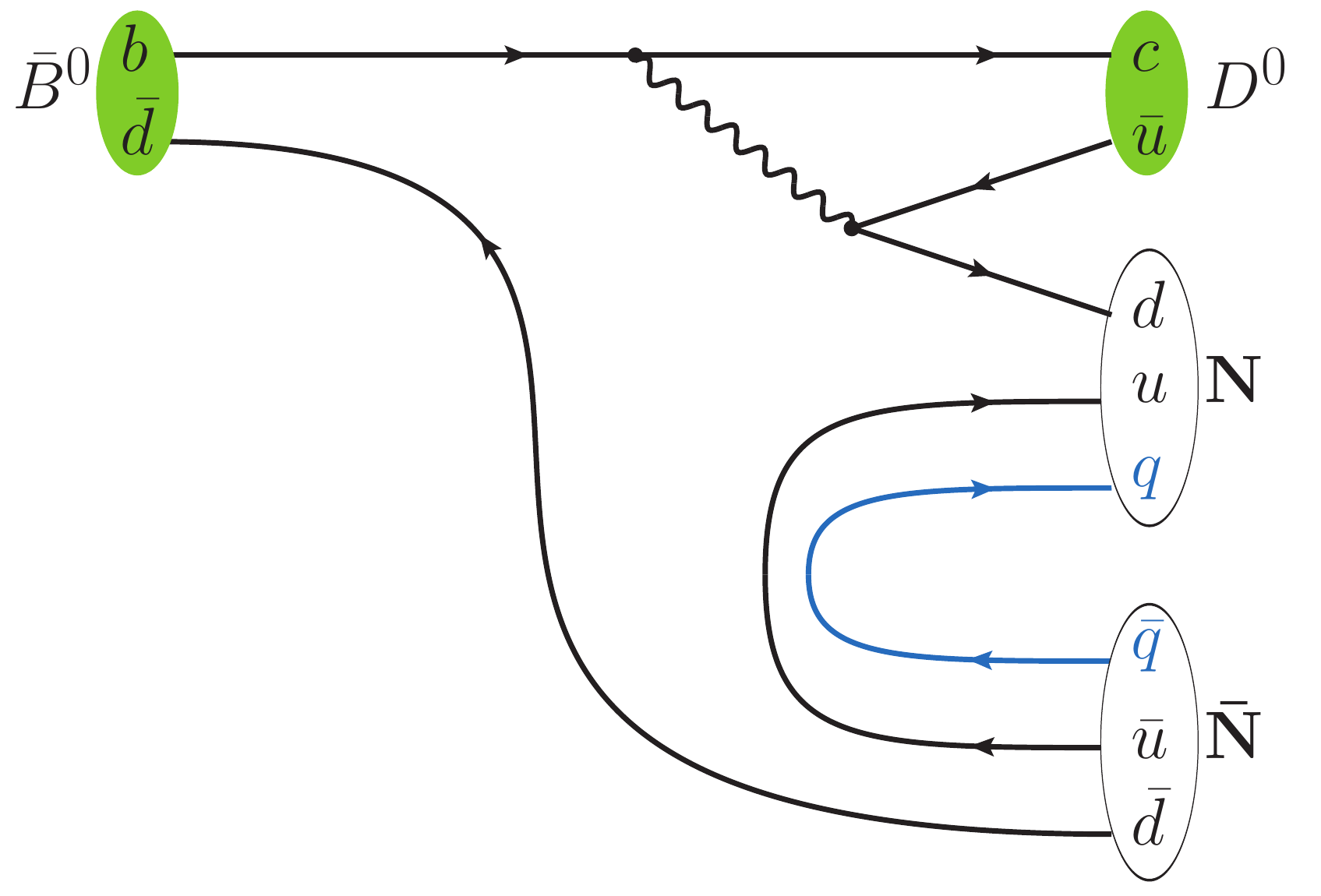}
   \includegraphics[width=.33\textwidth]{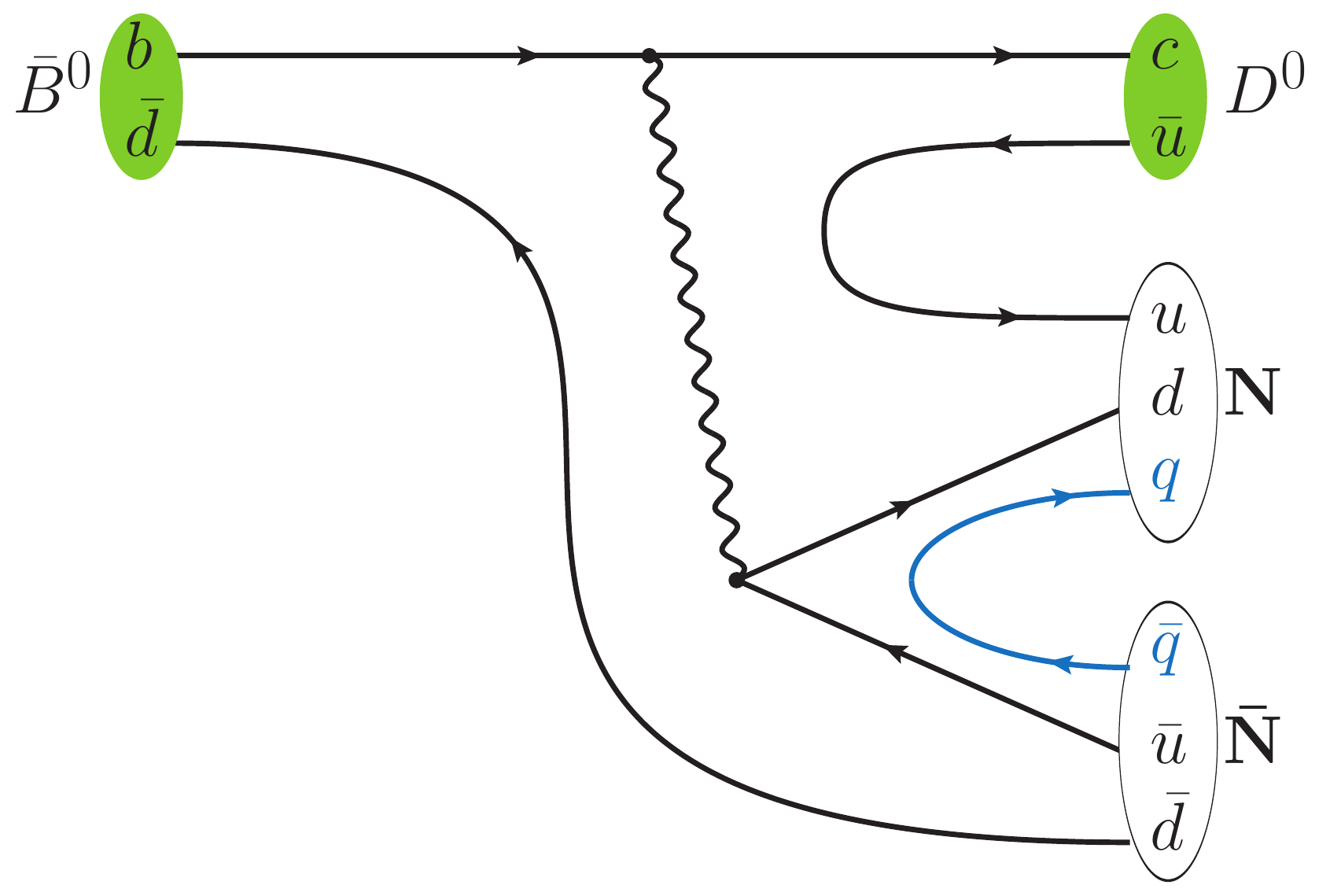}
   \includegraphics[width=.30\textwidth]{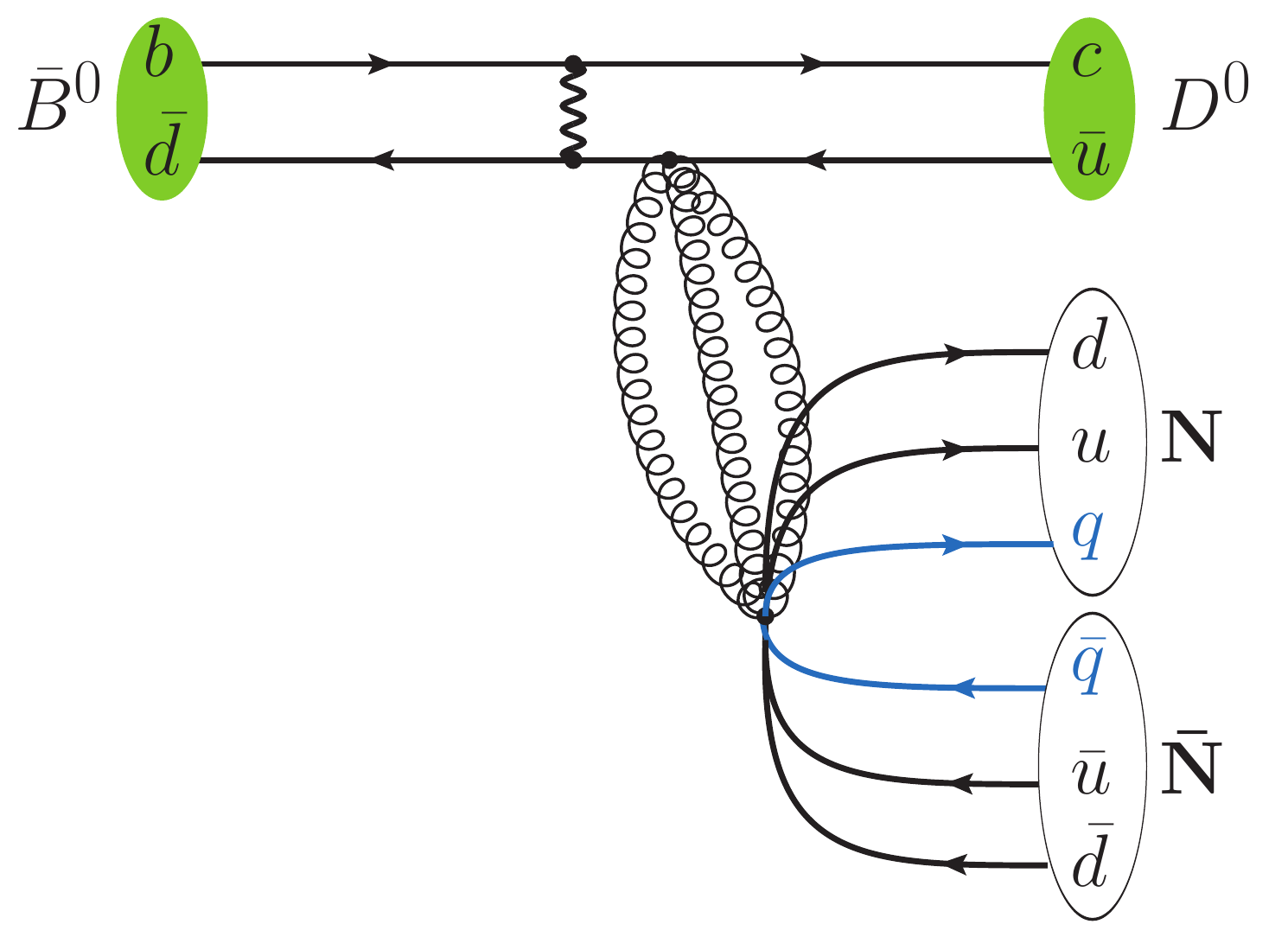}
 \caption{Leading-order Feynman diagrams for the decays $\Bzb\to\Dz\NN$. Setting $q=u$ leads to the $\Dz\pp$ final state and setting $q=s$ to the $\Dz\LL$, $\Dz\Sigmaz\Lambdabar$,
 $\Dz\Lambdaz\Sigmazbar$, and $\Dz\Sigmaz\Sigmazbar$ final states.}
 \label{fig:Feyn}
\end{figure*}

The decay of a \B meson into a \Dz meson and a pair of baryons has been the object of several theoretical investigations~\cite{ref:Cheng, ref:Hsiao}.
Ref.~\cite{ref:Hsiao} predicts the branching fractions for $\BDLL$  decays and for the sum of the $\BDLS$ and $\BDSL$ decays to be
\begin{equation}
 \label{eq:6}
 \begin{split}
   \BR(\BDLL) &= (2\pm1)\times 10^{-6} ,\\
   \BR(\BDLS+\BDSL)&=(1.8\pm0.5)\times 10^{-5} .
 \end{split}
\end{equation}
It is impractical to separate the $\BDLS$ and $\BDSL$ decays since each leads to the final state $\LL\g$.

As can be seen from the Feynman diagrams shown in Fig.~\ref{fig:Feyn}, the only difference between the \BDpp decay on the one hand and the \BDLL and \BDSL decays on the other hand is the replacement of a \uubar pair with an \ssbar pair.
In the hadronization process, \ssbar-pair production is suppressed by about a factor of three compared to \uubar- or \ddbar-pair production~\cite{ref:Lund}. 
Furthermore, since both \Lambdaz and \Sigmaz baryons can be produced, there are four possible final states with an \ssbar pair ($\LL$, $\Lambdaz\Sigmazbar$, $\Sigmaz\Lambdabar$, and $\Sigmaz\Sigmazbar$) compared to only one for a \uubar pair (\pp), neglecting the production of excited baryons.
Assuming equal production rates for these four modes and that the spin-$1/2$ states dominate, a suppression of a factor of $\mathord{\sim}12$ is expected for \BDLL decays compared to \BDpp decays, where the branching fraction of the latter process is $\BR(\BDpp) = \left( 1.04 \pm 0.04 \right) \times 10^{-4}$~\cite{ref:PDG}.

The branching fraction for \BDLL has been measured by the Belle Collaboration to be $\BR(\BDLL) = (10.5^{+5.7}_{-4.4} \pm 1.4)\times 10^{-6}$ \cite{ref:BelleDLL}. There are no previous results for the \BDSL decay mode.

\section{The \babar experiment}
\label{sec:babar-experiment}
This analysis is based on a data sample of $429 \invfb$ \cite{ref:Lumi}, corresponding to $471 \times 10^6$ \BBb pairs, collected with the \babar detector at the \pep2 asymmetric-energy \epem collider at the SLAC National Accelerator Laboratory at center-of-mass energies near and equal to the \FourS mass. 
The reconstruction efficiency is determined through use of Monte Carlo (MC) simulation, based on the EvtGen \cite{ref:EvtGen} program for the event generation and the GEANT4 \cite{ref:geant} package for modeling of the detector response. The MC events are generated uniformly in the $\BDLL$ and $\BDSL$ phase space. 

The \babar detector is described in detail elsewhere \cite{ref:NIM1, ref:NIM2}. 
Charged particle trajectories are measured with a five-layer double-sided silicon vertex tracker and a 40-layer drift chamber immersed in a $1.5$ T axial magnetic field. Charged particle identification is provided by ionization energy measurements in the tracking chambers and by  Cherenkov-radiation photons recorded with an internally reflecting ring-imaging detector. Electrons and photons are reconstructed with an electromagnetic calorimeter.

\section{Reconstruction of $\boldsymbol{\Lambdaz}$ baryon, $\boldsymbol{\Dz}$ meson, and $\boldsymbol{\Bzb}$ meson candidates}
\label{sec:Analysis}
We reconstruct \Lambdaz baryons through the decay mode $\Lambdaz\ra\proton\pim$ and \Dz mesons through the modes $\Dz\to\Km\pip$, $\Dz\to\Km\pip\pipi$, and $\Dz\ra\Km\pip\piz$~\cite{footnote}. Charged kaon and proton candidates are required to satisfy particle identification criteria. Charged pions are selected as charged tracks that are not identified as a kaon or proton.

Candidate \piz mesons are reconstructed from two separated energy deposits in the electromagnetic calorimeter not associated with charged tracks. To discriminate against neutral hadrons, the shower shape of each deposit is required to be consistent with that of a photon \cite{ref:LAT}. Furthermore, we require $E(\gamma_1) > 0.125\gev$ and $E(\gamma_2)>0.04\gev$, where $E(\g_1)$ and $E(\g_2)$ are the energies of the photon candidates, with $E(\g_1)>E(\g_2)$. The photon-photon invariant mass is required to lie in the range $m(\gaga)\in [0.116,0.145]\gevcc$.

The \Lambdaz daughters are fit to a common vertex and the reconstructed mass is required to lie within three standard deviations of the nominal value~\cite{ref:PDG}, where the standard deviation is the mass resolution. We select \Lambdaz candidates by requiring the flight significance $L_{t}/\sigma_{L_{t}}$ to exceed $4$, where $L_{t}$ is the \Lambdaz flight length in the transverse plane and $\sigma_{L_{t}}$ its uncertainty. The \Sigmaz baryons are produced in the decay $\Sigmaz\to\Lambdaz\g$, and the photon is not reconstructed.

\begin{figure}[t]
  \centering
  \includegraphics[width=.9\columnwidth]{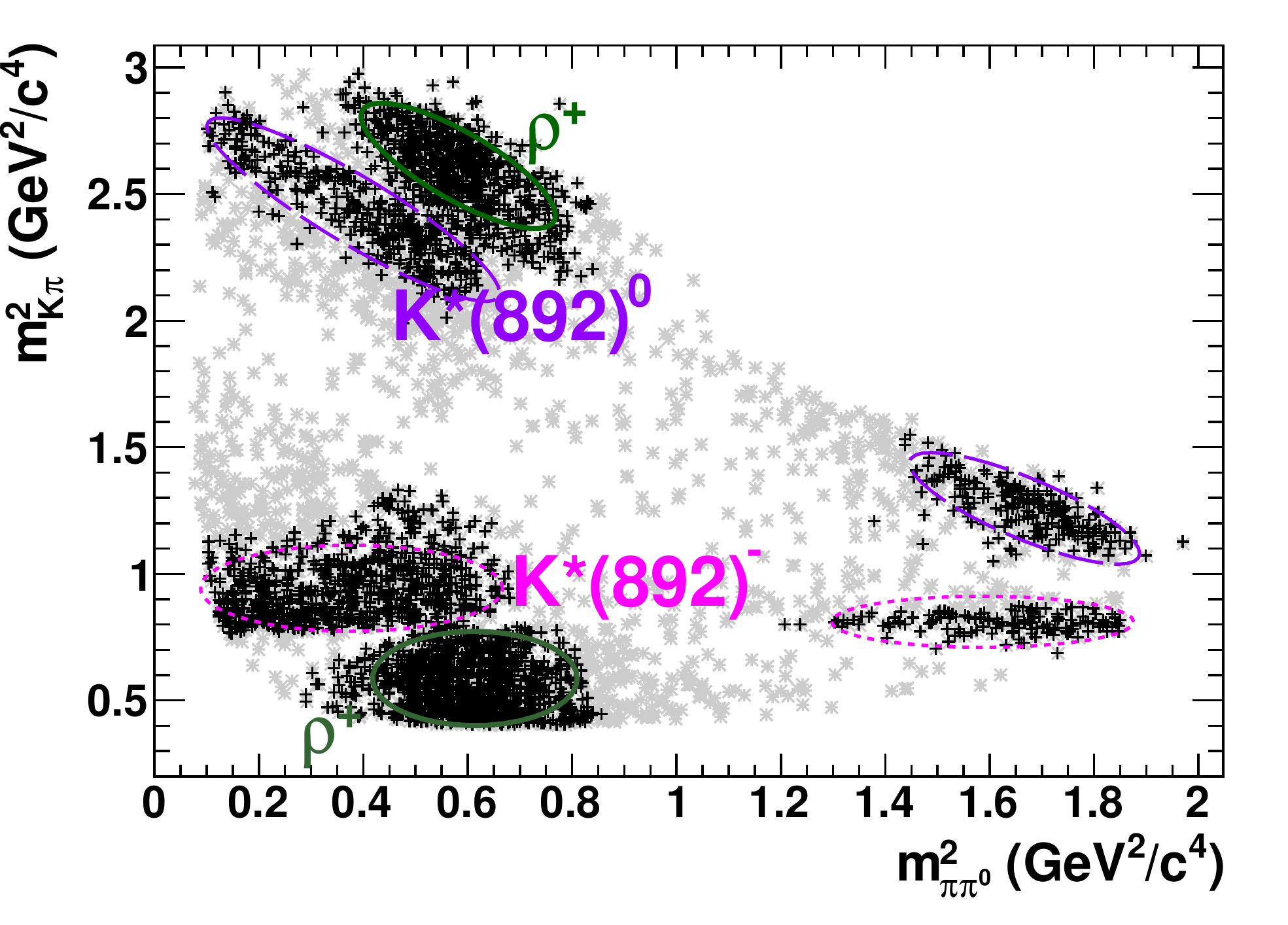}
  \caption{Dalitz plot for simulated $\Dz\to\Km\pip\piz$ events before (gray stars) and after (black crosses) the $w_{\rm Dalitz}>0.02$ requirement. Resonant decays are indicated.}
  \label{fig:Dalitz}
\end{figure}

The \Dz daughter candidates are fit to a common vertex and the reconstructed mass is required to lie within three times the mass resolution from their nominal values~\cite{ref:PDG}. 
The signal-to-background ratio for $\Dz\to\Km\pip\piz$ is improved by making use  of the resonant substructure of this decay, which is well known. Using results from the E691 Collaboration \cite{ref:weights}, we calculate the probability $w_{\rm Dalitz}$ for a \Dz candidate to be located at a certain position in the Dalitz plane. 
We require $w_{\rm Dalitz} > 0.02$. 
Figure~\ref{fig:Dalitz} shows the Dalitz plot distributions, based on simulation, for candidates selected with and without the $w_{\rm Dalitz}$ requirement.

\begin{figure*}[!t]
  \centering
  \includegraphics[width=1.0\textwidth]{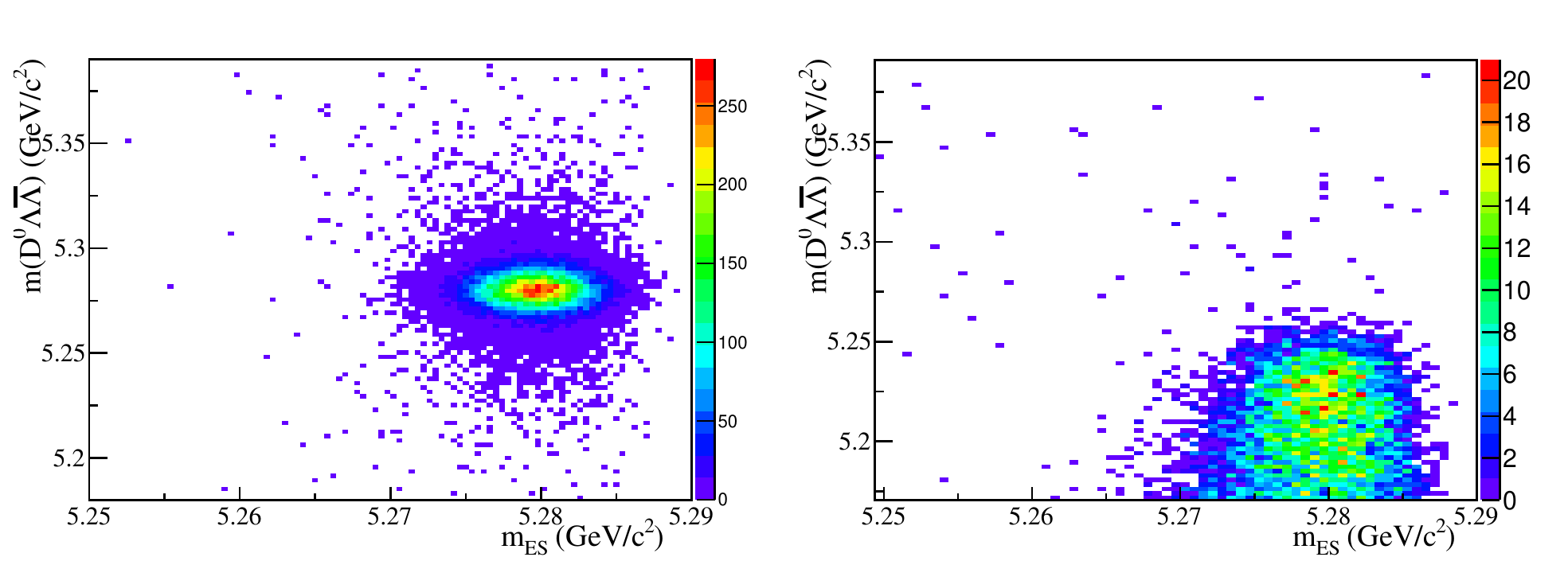}
  \caption{Distributions for \BDLL (left) and  \BDSL  reconstructed as \BDLL (right) for the $\Dz\to\Km\pip$ mode in simulated events.}
  \label{fig:SigmaVergl}
\end{figure*}

The \Dz and \Lambdaz candidates are constrained to their nominal masses in the reconstruction of the \Bzb candidates. We apply a fit to the entire decay chain and require the probability for the vertex fit to be larger than $0.001$.

To reduce background from $\epem\to\qqbar$ events with $\q = \u, \d, \s, \c$, we apply a selection on a Fisher discriminant $\mathcal{F}$ that combines the values of $\left|\cos\theta_{\rm Thr}\right|$, where $\theta_{\rm Thr}$ is 
the angle between the thrust axis of the \B candidate and the thrust axis formed from the remaining tracks and clusters in the event; $\left|\cos\theta_{z}\right|$, where $\theta_{z}$  is the angle between the $\B$ thrust axis and the beam axis; $\left|\cos\phi\right|$, where $\phi$ is the angle between the $\B$ momentum and the beam axis; and the normalized second Fox Wolfram moment \cite{ref:FoxWolf}. All these quantities are defined in the center-of-mass frame. All selection criteria are summarized in Table~\ref{tab:Cuts}.

\begin{table}[!h]
  \centering
  \caption{Summary of selection criteria.}
  \begin{tabular}{l|c}
    \hline
    Selection criterion           & Selected candidates                       \\
    \hline
    $\Lambdaz / \Lambdabar$ mass &  $m_{\proton\pion}\in [1.112,1.120]\gevcc$   \\
    Flight significance          &  $L_{t}/\sigma_{L_{t}}>4$                  \\
    \hline
    $\Dz\to\Km\pi$ mass          &  $m_{\kaon\pion}\in [1.846,1.882]\gevcc$  \\
    \hline
    $\Dz\to\Km\pip\pipi$ mass    &  $m_{\kaon\pion\pion\pion}\in [1.852,1.876]\gevcc$ \\
    \hline
    Lateral parameter $\g_1$     &  $0.05<\mathrm{LAT}(\g_1)<0.55$                       \\
    Lateral parameter $\g_2$     &  $\mathrm{LAT}(\g_2)>0.075$                       \\
    Calorimeter energy $\g_1$    &  $E(\g_1)>0.125\gev$        \\
    Calorimeter energy $\g_2$    &  $E(\g_2)>0.04\gev$          \\
    \piz mass                    &  $m_{\gaga}\in [0.116,0.145]\gevcc$        \\
    $\Dz\to\Km\pip\piz$ mass      &  $m_{\kaon\pion\piz}\in [1.81,1.89]\gevcc$  \\
    {Dalitz} weight       &  $\DW>0.02$                              \\
    \hline
    \B vertex probability        &  $p(\B)>0.001$                            \\
    Fisher discriminant          &  $\mathcal{F}>0.1$                         \\
    \hline
  \end{tabular}
  \label{tab:Cuts}
\end{table}

\section{Fit strategy}
\label{sec:fit-strategy} 

We determine the number of signal candidates with a two-dimensional unbinned extended maximum likelihood fit to the invariant mass $\minv$ and the energy substituted mass \mes. The latter is defined as
\begin{equation}
  \label{eq:7}
  \mes = \sqrt{(s/2+\boldsymbol{p}_0 \cdot \boldsymbol{p}_{\B})^2/E_0^2-\left|\boldsymbol{p}\right|_{\B}^2} ~,  
\end{equation}
where $\sqrt{s}$ is the center-of-mass energy, $\boldsymbol{p}_B$ the \B candidate's momentum, and $(E_0,\boldsymbol{p}_0)$ the four-momentum vector of the \epem system, each given in the laboratory frame. Both \minv and \mes are centered at the \B mass for well reconstructed \B decays.

Due to the small mass difference of $76.9\mevcc$ \cite{ref:PDG} between the \Lambdaz and \Sigmaz baryons,
\BDSL decays, where the \Sigmaz decays radiatively as $\Sigmaz\to\Lambdaz\g$, are a source of background. Such events peak at the \B mass in \mes and are slightly shifted in \minv with respect to \BDLL (Fig.~\ref{fig:SigmaVergl}).
We account for this decay by including an explicit term in the likelihood function (see below), whose yield is determined in the fit.

We divide the data sample into three subsamples corresponding to the \Dz decay modes. Given their different signal-to-background ratios, we determine the number of signal candidates in a simultaneous fit to the three independent subsamples.
We describe each \BDLL signal sample with the product of a Novosibirsk function in \mes and a sum of two Gaussian functions $f^{\mathcal{GG}}$ in \minv assuming that \mes and \minv are not correlated. We study simulated samples of signal and background events and find no significant correlation between \mes and \minv. The Novosibirsk function is defined as
 \begin{equation}
   \label{eq:1}
   \begin{split}
     f^{\rm Novo}(\mes) &= \exp \left[ -\frac{1}{2} \left( \frac{ \ln^2 [1 + \lambda \alpha (\mes - \mu) ]} { \alpha^2} + \alpha^2 \right)  \right]\ , \\
     \lambda &= {\sinh ( \alpha \sqrt{\ln 4} )} / {( \sigma \alpha \sqrt{\ln 4} )} ,
   \end{split}
 \end{equation}
with $\mu$ the mean value, $\sigma$ the width, and $\alpha$ the tail parameter.
The decay \BDSL is described by the product of a Novosibirsk $f^{{\rm Novo}1,\Sigmaz}$ function in \mes and a sum of another Novosibirsk function $f^{{\rm Novo}2,\Sigmaz}$ and a Gaussian $\mathcal{G}^{\Sigmaz}$ in \minv. All parameters are determined using Monte Carlo simulated events and are fixed in the final fit. Background from $\epem\to\qqbar$ events and other \B meson decays is modeled by the product of an ARGUS function \cite{ref:argus} in \mes and a first order polynomial in \minv.

The full fit function is defined as
\begin{widetext}
  \begin{equation}
    \label{eq:8}
    \begin{split}
      f^{\rm Fit}_j & = f^{\Lambdaz}_j + f^{\Sigmaz}_j + f^{\rm Bkg}_j\\ 
      &= f^{{\rm Novo},\Lambda}_j(\mes)\times f^{\mathcal{GG}}_j\left(\minv\right) + f^{{\rm Novo}1,\Sigmaz}_j(\mes)\times \left[f^{{\rm Novo}2,\Sigmaz}_j\left(\minv\right) + \mathcal{G}^{\Sigmaz}_j\left(\minv\right)\right] \\
      &+ f^{\rm ARGUS}_j(\mes) \times f^{\rm Poly}_j\left(\minv\right) 
      ,
    \end{split}
  \end{equation}
\end{widetext}
where the index $j$ corresponds to the three \Dz decay modes.

The branching fraction is determined from
\begin{equation}
  \begin{split}
  \BR(\BDLL) &= \frac{N(\BDLL)}{2N_{\BzBzbar} \times \varepsilon^\Lambdaz} \\
  &\times \frac{1}{\BR(\Lambdaz\to\proton\pion)^2\BR(\Dz\to X)} ,
  \end{split}
\end{equation}
where $N(\BDLL)$ is the fitted signal yield, $N_{\BzBzbar}$ the number of the \BzBzbar pairs assuming $\BR(\FourS\to\BzBzbar)=0.5$, $\varepsilon^\Lambda$ the total reconstruction efficiency, and $\BR(\Lambdaz\to\proton\pion)$ and $\BR(\Dz\to X)$ the branching fractions for the daughter decays of \Lambdaz and \Dz, respectively. 
An analogous expression holds for $\BR(\BDSL)$. We perform a simultaneous fit of the three \Dz decay channels to obtain:
\begin{equation}
  \begin{split}
    N_\Lambdaz &= \frac{N(\BDLL)}{\varepsilon^\Lambdaz\BR(\Dz\to X)} , \\
    N_{\Sigmaz} &= \frac{N(\BDSL)}{\varepsilon^{\Sigmaz}\BR(\Dz\to X)} .
  \end{split}
\end{equation}

The likelihood function is given by
\begin{widetext}
 \begin{equation}
   \begin{split}
     \label{eq:like3}
     L &= \prod_j\frac{e^{-\left( \varepsilon^{\Lambdaz}_j \BR_jN_{{\Lambdaz}} + N^{\mathrm{Bkg}}_{j} + \varepsilon_j^{\Sigmaz} \BR_j N_{\Sigmaz}\right)}}{N(j)!} \prod_k^{N(j)} \left[ \varepsilon_j^{\Lambdaz} \BR_jN_{{\Lambdaz}} f^{\Lambdaz}_j\left(\mes_k, {\minv}_k\right) \right. 
     +  N^{\mathrm{Bkg}}_{j} f^{\mathrm{Bkg}}_{j}\left(\mes_k,{\minv}_k\right) \\ & \left.+ \varepsilon_j^{{\Sigmaz}} \BR_jN_{{\Sigmaz}} f^{{\Sigmaz}}_j\left(\mes_k, {\minv}_k\right)\right] ,
   \end{split}
 \end{equation}
\end{widetext}
where $\BR_j$ is the branching fraction for the $j$th \Dz decay, $N^{\mathrm{Bkg}}_{j}$ the number of combinatorial background events in the $j$th subsample, $N_{\Lambdaz}$ and $N_{\Sigmaz}$ the yields of \BDLL and \BDSL, and $\varepsilon^{\Lambdaz}_j$ and $\varepsilon_j^{\mathrm{\Sigmaz}}$ the efficiencies for the $j$th \Dz decay.

\section{Systematic uncertainties}
\label{sec:syst-uncert}
We consider the following systematic uncertainties: the uncertainties associated with the number of \BBbar events, the particle identification (PID) algorithm, the tracking algorithm, the \piz reconstruction, the \Dz and \Lambdaz branching fractions, the efficiency correction, and the fitting algorithm.

The uncertainty associated with the number of \BBbar pairs is $0.6\%$. 
We determine the systematic uncertainty associated with the PID by applying different PID selections and comparing the result with the nominal selection.
The difference is $0.8\%$, which is assigned as the PID uncertainty. 
The systematic uncertainty associated with the tracking algorithm depends on the number of charged tracks in the decay. We assign a systematic uncertainty of $0.9\%$ for the $\Dz\to\Km\pip$ and $\Dz\to\Km\pip\piz$ decays and $1.2\%$ for the $\Dz\to\Km\pip\pipi$ decay. A $3\%$ uncertainty is assigned to account for the \piz reconstruction in $\Dz\to\Km\pip\piz$ decays. A detailed description of these detector-related systematic uncertainties is given in Ref.~\cite{ref:NIM2}.

We rely on the known \Dz branching fractions in our fit. To estimate the associated systematic uncertainty we vary each branching fraction by one standard deviation of its uncertainty~\cite{ref:PDG} and define the systematic uncertainty to be the maximum deviation observed with respect to the nominal analysis. 
We divide $m(\LL)$ into six bins and determine the total reconstruction efficiency $\varepsilon_i$ in each bin. We determine the uncertainty due to the use of the average efficiency $\bar{\varepsilon}$ by studying $|\varepsilon_i - \bar{\varepsilon}|/\bar{\varepsilon}$ as a function of $m(\LL)$. We average these values and take the result of $16.3\%$ ($\Dz\to\Km\pip$), $19.6\%$ ($\Dz\to\Km\pip\piz$), and $16.8\%$ ($\Dz\to\Km\pip\pipi$) as our estimate of the systematic uncertainty for the efficiency.
We estimate the systematic uncertainty due to the fit procedure by independently varying the fit ranges of \mes and \minv. 
The largest differences in the signal yield are  $3.9\%$ for the change of the \mes fit range and $2.1\%$ for the change of the \minv fit range. To check our background model, we use a second-order polynomial in \minv instead of a first-order polynomial. The signal yield changes by $1.1\%$.
We use an ensemble of simulated data samples reflecting our fit results to verify the stability of the fit. We generate 1000 such samples with shapes and yields fixed to our results and repeat the final fit. We find no bias in the signal-yield results.
All systematic uncertainties are summarized in Table \ref{tab:syserr}.

The total systematic uncertainty, obtained by adding all sources in quadrature, is $20.1\%$.

\begin{table}[h]
  \centering
  \caption{Summary of the systematic uncertainties for \BDLL.}
  \begin{tabular}{l|r}
    \toprule
    Source          & Relative uncertainty \\
    \hline
    \BBbar counting & $0.6\%$                \\
    Particle identification    & $0.8\%$                \\
    Tracking        &                  \\
    \quad $\Dz\to\Km\pip$ & $0.9\%$ \\
    \quad $\Dz\to\Km\pip\piz$ & $0.9\%$ \\
    \quad $\Dz\to\Km\pip\pipi$ & $1.2\%$ \\
    \piz systematics & \\
    \quad $\Dz\to\Km\pip\piz$ & $3.0\%$ \\
    \Dz and \Lambdaz branching fractions       & $2.9\%$                  \\
    Efficiency correction&            \\
    \quad $\Dz\to\Km\pip$ & $16.3\%$ \\
    \quad $\Dz\to\Km\pip\piz$ & $19.6\%$ \\
    \quad $\Dz\to\Km\pip\pipi$ & $16.8\%$ \\
    Fit procedure             &   $4.6\%$               \\
    \hline
    Total uncertainty          &   $20.1\%$              \\
    \hline
    \hline
  \end{tabular}
  \label{tab:syserr}
\end{table}

\section{Results}
\label{sec:dynamic-structure}
\begin{figure*}[t]
  \centering
  \includegraphics[width=1.\textwidth]{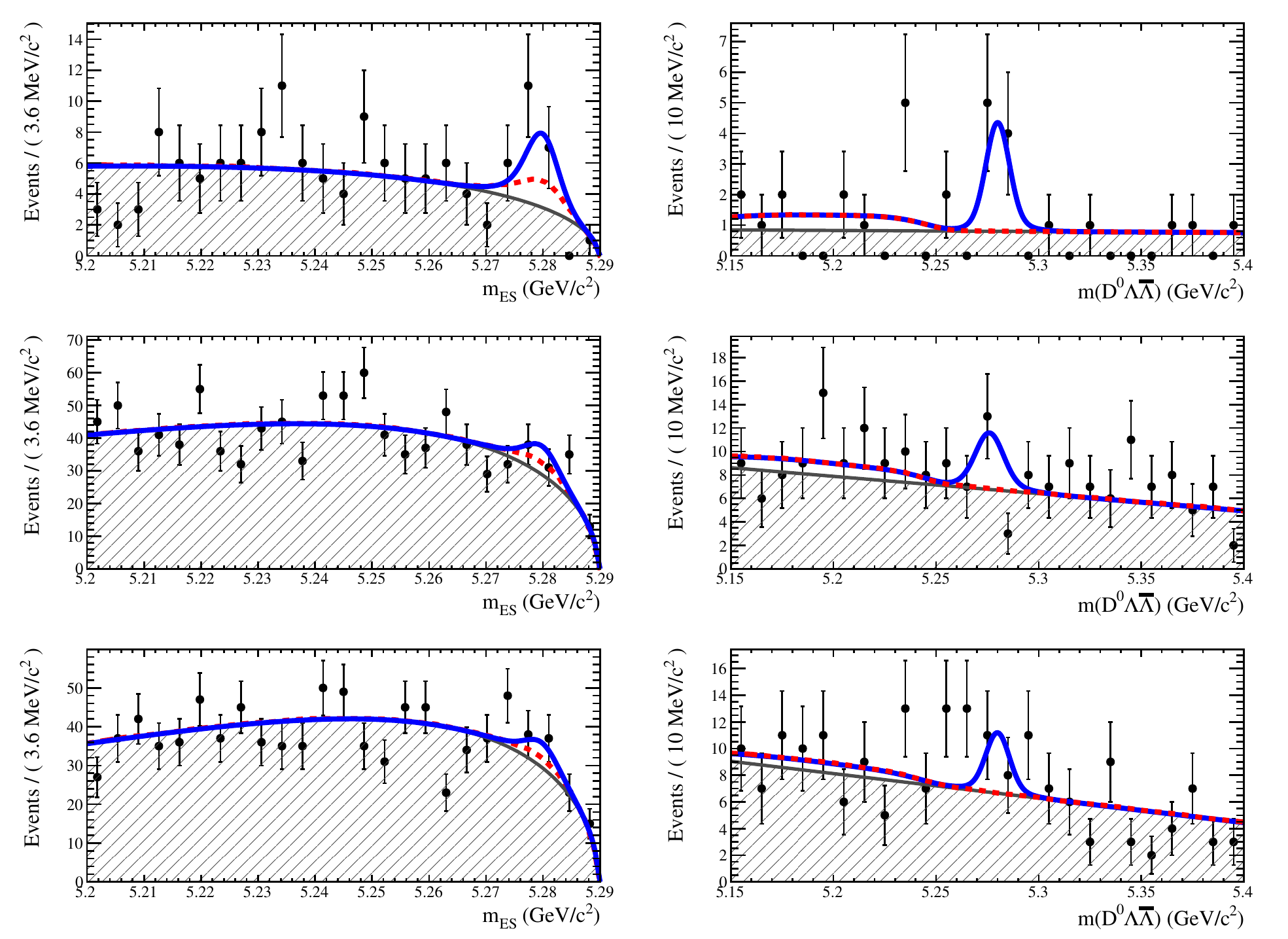}
  \caption{Results of the combined fit. The \mes projection is shown for $\minv\in[5.15,5.31]\gevcc$ and the \minv projection for $\mes\in[5.272,5.286]\gevcc$. The solid line shows the result of the fit, the dashed curve indicates the $\Bzb\to\Dz\Sigmaz\Lambdabar$ contribution, and the shaded histogram the combinatorial background. From top to bottom: $\Dz\to\Km\pip$, $\Dz\to\Km\pip\piz$, and $\Dz\to\Km\pip\pipi$ subsamples.}
  \label{fig:Datenfit}
\end{figure*}

The one-dimensional projections of the fit are shown in Fig.~\ref{fig:Datenfit}. We find 
\begin{equation}
  \label{eq:5}
  \begin{split}
    N_{\Lambdaz} &= 1880^{+560}_{-500} , \\
    N_{\Sigmaz} &= 2870^{+1680}_{-1560} .    
  \end{split}
\end{equation}
The statistical significance is calculated as $\sqrt{-2\log{L_0/L_S}}$, where $L_0$ is the likelihood value
for a fit without a signal component and $L_S$ is the likelihood value
for the nominal fit. The statistical significance of the combined \BDLL and \BDSL yields is $3.9$ standard deviations ($\sigma$), while those of the individual \BDLL  and \BDSL results are $3.4\sigma$ and $1.2\sigma$, respectively. Compared to the statistical uncertainty the additive systematic uncertainties are negligible. We therefore quote the statistical significance as the global significance.

The branching fractions are
\begin{equation}
  \label{eq:2}
  \begin{split}
    &\BR(\BDLL) =\left( 9.8^{+2.9}_{-2.6} \pm 1.9\right) \times 10^{-6} ,\\
    &\BR(\BDSL+\BDLS) =\\ & \left( 15^{+9}_{-8} \pm 3 \right) \times 10^{-6} ,\\
  \end{split}
\end{equation}
where the first uncertainties represent the statistical uncertainties and the second the systematic uncertainties.
As a cross-check of the method, independent fits to the three sub-samples are performed.
The results of each of these fits are consistent with each other and with the nominal combined fit.

Since the statistical significance for $\BR(\BDSL+\BDLS)$ is low, a Bayesian upper limit at the $90\%$ confidence level is calculated by integrating the likelihood function:
\begin{equation}
  \label{eq:UL}
  \BR(\BDSL+\BDLS)<3.1\times 10^{-5} .
\end{equation}

\begin{figure}[t]
  \centering
  \includegraphics[width=.9\columnwidth]{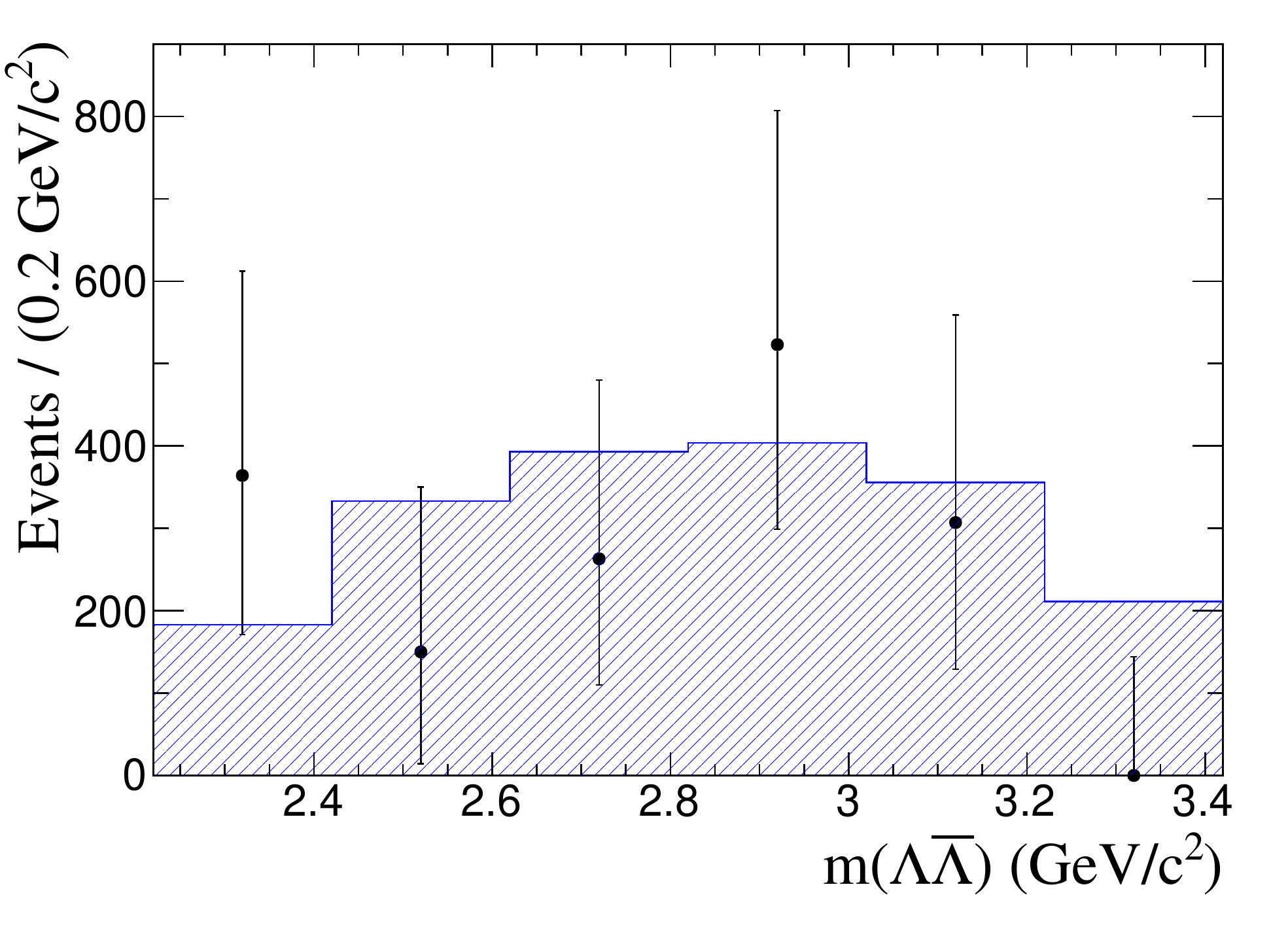}
  \caption{Distribution of the invariant baryon-antibaryon mass for \Dz-branching-fraction and efficiency-corrected \BDLL signal candidates. The data points represent the \babar data and the shaded histogram indicates phase-space-distributed simulated events, scaled to match the area under the data.}
  \label{fig:llmassspectrum}
\end{figure}

To investigate the threshold dependence, we perform the fit in bins of $m(\LL)$ and examine the resulting distribution after accounting for the reconstruction efficiency and $\Dz$ branching fractions. The results are shown in Fig.~\ref{fig:llmassspectrum}. No enhancement in the $\BDLL$ event rate is observed at the baryon-antibaryon mass threshold within the uncertainties, in contrast to $\BDpp$ decays, which do exhibit such an enhancement \cite{ref:Dpp}.

We compare our results for the $\BDLL$ and $\BDSL$ branching fractions to theoretical predictions. The result we obtain for the $\BDSL$ branching fraction is consistent with the prediction of $\BR(\BDSL+\BDLS) = \left( 18\pm 5\right) \times 10^{-6}$ from Ref. \cite{ref:Hsiao}.
However, the obtained result for the $\BDLL$ branching fraction is larger than the prediction of  $\BR(\BDLL) = (2\pm1)\times 10^{-6}$ \cite{ref:Hsiao} by a factor of
\begin{equation}
  \label{eq:3}
  \frac{\BR(\BDLL)_\mathrm{exp}}{\BR(\BDLL)_\mathrm{theo}} = 4.9 \pm 3.0 .
\end{equation}

We further determine
\begin{equation}
  \label{eq:9}
  \frac{\BR(\BDSL + \Bzb\to\Dz\Lambdaz\Sigmazbar)}{\BR(\BDLL)} = 1.5\pm 0.9 ,
\end{equation}
which is in agreement with our assumption that all four modes \BDLL, \BDSL,
  $\Bzb\to\Dz\Lambdaz\Sigmazbar$, and $\Bzb\to\Dz\Sigmaz\Sigmazbar$ are produced at equal rates.
For the ratio of branching fractions we find 
\begin{equation}
  \label{eq:4}
  \frac{\BR(\BDLL)}{\BR(\BDpp)} = \frac{1}{10.6\pm3.7} ,
\end{equation}
using $\BR(\BDpp) = (1.04\pm0.04) \times 10^{-4}$~\cite{ref:PDG}.
This is in agreement with the expected suppression of $1/12$ discussed in the introduction.

\section{Summary}
\label{sec:conclusions}
We find evidence for the baryonic \B decay \BDLL. We determine the branching fraction to be $\BR(\BDLL) = (9.8^{+2.9}_{-2.6} \pm 1.9)\times 10^{-6}$ with a significance of $3.4\sigma$ including systematic uncertainties. This is in agreement with the Belle measurement \cite{ref:BelleDLL}. We find no evidence for an enhancement in the invariant baryon-antibaryon mass distribution near threshold. Our result for the branching fraction deviates from theoretical predictions based on measurements of \BDpp but agrees with simple models of hadronization.
We find no evidence for the decay \BDSL and calculate a Bayesian upper limit at $90\%$ confidence level of $\BR(\BDSL+\BDLS) < 3.1 \times 10^{-5}$. This result is in agreement with the theoretical expectation.

We are grateful for the 
extraordinary contributions of our \pep2\ colleagues in
achieving the excellent luminosity and machine conditions
that have made this work possible.
The success of this project also relies critically on the 
expertise and dedication of the computing organizations that 
support \babar.
The collaborating institutions wish to thank 
SLAC for its support and the kind hospitality extended to them. 
This work is supported by the
US Department of Energy
and National Science Foundation, the
Natural Sciences and Engineering Research Council (Canada),
the Commissariat \`a l'Energie Atomique and
Institut National de Physique Nucl\'eaire et de Physique des Particules
(France), the
Bundesministerium f\"ur Bildung und Forschung and
Deutsche Forschungsgemeinschaft
(Germany), the
Istituto Nazionale di Fisica Nucleare (Italy),
the Foundation for Fundamental Research on Matter (The Netherlands),
the Research Council of Norway, the
Ministry of Education and Science of the Russian Federation, 
Ministerio de Ciencia e Innovaci\'on (Spain), and the
Science and Technology Facilities Council (United Kingdom).
Individuals have received support from 
the Marie-Curie IEF program (European Union), the A. P. Sloan Foundation (USA) 
and the Binational Science Foundation (USA-Israel).

\end{document}